\documentclass[apsl,prx,twocolumn,groupedaddress,showpacs,showkeys]{revtex4-1}

\usepackage{graphicx}
\usepackage{dcolumn}
\usepackage{bm}

\begin{document}

\title{
Atomistic Mechanisms of Codoping-Induced p- to n-Type Conversion in Nitrogen-Doped Graphene}

\author{Hyo Seok Kim}
\affiliation{Graduate School of Energy, Environment, Water, and Sustainability, Korea Advanced Institute of Science and Technology, 291 Daehak-ro, Yuseong-gu, Daejeon 305-701, Korea}
\author{Han Seul Kim}
\affiliation{Graduate School of Energy, Environment, Water, and Sustainability, Korea Advanced Institute of Science and Technology, 291 Daehak-ro, Yuseong-gu, Daejeon 305-701, Korea}
\author{Seong Sik Kim}
\affiliation{Graduate School of Energy, Environment, Water, and Sustainability, Korea Advanced Institute of Science and Technology, 291 Daehak-ro, Yuseong-gu, Daejeon 305-701, Korea}
\author{Yong-Hoon Kim}
 \email[Corresponding author.\\]{y.h.kim@kaist.ac.kr}
\affiliation{Graduate School of Energy, Environment, Water, and Sustainability, Korea Advanced Institute of Science and Technology, 291 Daehak-ro, Yuseong-gu, Daejeon 305-701, Korea}

\date{\today}

\begin{abstract}
It was recently shown that nitrogen-doped graphene (NG) can exhibit both p- and n-type characters depending on the C-N bonding nature, which represents a significant bottleneck for the development of graphene-based electronics. Based on first-principles calculations, we herein scrutinise the correlations between atomic and electronic structures of NG, and particularly explore the feasibility of converting p-type NG with pyridinic, pyrrolic, and nitrilic N into n- or bipolar type by introducing an additional dopant atom. Out of the nine candidates, B, C, O, F, Al, Si, P, S, and Cl, we find that the B-, Al-, and P-codoping  can anneal even relatively large vacancy defects in p-type NG. It will be also shown that, while the NG with pyridinic N can be converted into n-type via codoping, only the bipolar type conversion can be achieved for the NG with nitrilic or pyrrolic N. The amount of work function reduction was up to 0.64 eV for the pyridinic N next to a monovacancy. The atomistic origin of such diverse type changes is analysed based on Mulliken and crystal orbital Hamiltonian populations, which provide us with a framework to connect the local bonding chemistry with macroscopic electronic structure in doped and/or defective graphene. Moreover, we demonstrate that the proposed codoping scheme can recover the excellent charge transport properties of pristine graphene. Both the type conversion and conductance recovery in codoped NG should have significant implications for the electronic and energy device applications.
\end{abstract}


\maketitle

\section{Introduction\label{sec:intro}}

Because of its unique structural, electronic, and transport properties,
\cite{A.K.Geim_2013_1,A.H.CastroNeto_2009_2,N.M.R.Peres_2010_3,S.DasSarma_2011_4}
graphene is regarded as one of the best candidate materials to be incorporated into the next-generation electronic, energy, and bio devices.
\cite{K.Kim_2011_5,K.S.Novoselov_2012_6,J.Liu_2012_7}
To realize such potential for device applications, reliable methods to tailor the atomic and electronic structures of graphene is required.
\cite{H.Liu_2011_8,S.T.Pantelides_2012_9,U.N.Maiti_2014_10}
Doping of graphene has been extensively investigated in this context, and among various options nitrogen doping emerged as one of the most effective schemes to improve diverse functionalities of graphene
\cite{U.N.Maiti_2014_10,Y.Wang_2010_11,L.Qu_2010_12,A.L.M.Reddy_2010_13,H.Wang_2011_14,Z.Luo_2011_15,H.M.Jeong_2011_16,L.Lai_2012_17,H.Wang_2012_18,M.J.Ju_2013_19,X.Wang_2014_20} 
and especially achieve n-type doping that is crucial for electronics applications.
\cite{X.Wang_2009_21,B.Guo_2010_22,C.Zhang_2011_23,D.Usachov_2011_24,T.Schiros_2012_25}
Unfortunately, recent experimental and theoretical studies have revealed that N-doped graphene (NG) can assume both n- and p-type characters depending on the bonding nature of N atoms, namely, n-type for the graphitic N and p-type for the pyridinic, pyrrolic, and nitrilic Ns.
\cite{T.Schiros_2012_25,Y.Fujimoto_2012_26,Z.Hou_2013_27}
The question then arises as to how one can achieve robust n-type NG, and at the more fundamental level which factors determine the structure-property relationships of NG. 
The latter should have significant implications for practical applications. For example, the bonding state of the N atom was found to critically affect the oxygen reduction reaction (ORR) activity in fuel cell cathodes, but the nature of catalytically active sites remains controversial.
\cite{Z.Luo_2011_15,L.Lai_2012_17}

Carrying out first-principles density functional theory (DFT) and DFT-based non-equilibrium 
Green's function (NEGF) calculations, we herein systematically investigate the atomistic origins of p-type character in NG with pyridinic, pyrrolic, and nitrilic N and the feasibility of achieving robust n-type graphene by incorporating a codopant atom. Graphene codoped with N and B or N and P has been already synthesised and shown to improve the ORR performance,
\cite{S.Wang_2012_28,C.H.Choi_2013_29}
but the atomistic details and mechanism of the synergetic effects associated with the B, N- and P, N-codoping are not understood yet. We first consider the energetic feasibility of introducing an additional atom into various p-type NG defect sites and show that B, Al, Si and P atoms can structurally anneal even relatively large vacancy defects next to the pyridinic, nitrilic, and pyrrolic N atoms. We find that, except Si, the B, Al, and P codoping can convert the pyridinic NG into n-type and the nitrilic and pyrrolic NG into bipolar type, and thus effectively eliminate the p-type character of NG. Based on the Mulliken and crystal overlap Hamilton population (COHP) analyses, we further establish the basis to understand how the macroscopic electronic type change in NG is induced in the atomistic bonding viewpoint. Finally, we will demonstrate that the additional benefit of the codoping approach is the recovery of excellent charge transport capacity of pristine graphene, in line with the recovery of the $sp^{2}$ bonding network upon the healing of vacancy defects.

\section{Calculation Methods\label{sec:methods}} 

We performed the spin-polarized DFT calculations within the Perdew-Burke-Ernzerhof parameterization of generalized gradient approximation
\cite{J.P.Perdew_1996_30}
using the \textsf{SIESTA} software.
\cite{M.S.Jose_2002_31}
The atomic cores were replaced by the Troulier-Martins-type norm-conserving pseudopotentials,
\cite{N.Troullier_1991_32}
and the double-$\zeta$-plus-polarization level numerical atomic basis sets defined by the confinement energy of 80 meV were adopted. For the supercell of $9\times9$ graphene unit cells (see Supporting Information Fig. S1), we used a mesh cut-off energy of 200 Ry and the $3\times3\times1$ $\vec{k}$-point sampling in the Monkhost-Park scheme.
\cite{H.J.Monkhorst_1976_33} 
A finer $30\times30\times1$ $\vec{k}$ -point mesh was sampled for the calculation of density of states (DOS) and COHP.  
The vacuum region of the supercell along the direction perpendicular to the graphene plane was set to be 25 \AA.
We have calculated the work function $\Phi$ using the equation:
\begin{equation}
 \Phi = V_{vac} - E_F,
\end{equation}
where $E_F$ is the Fermi energy and $V_{vac}$ is the macroscopic average potential in the vacuum, defined as the midpoint between the graphene layer and its neighbouring images. 
\cite{D.Ziegler_2011}
The doping charge density was calculated according to 
\begin{equation}
n = \frac{(E_D - E_F)^2}{\pi (\hbar v_F)^2},
\end{equation}
where $E_D$ is the Dirac point energy and $v_F$ is the Fermi velocity, $1 \times 10^6$ m/s.
\cite{N.M.R.Peres_2010_3,D.Ziegler_2011}

Transmission functions $T(E)$ were calculated using the DFT-based NEGF method,
\cite{S.Datta_2013_34}
as implemented in
\textsf{TranSIESTA}.
\cite{M.Brandbyge_2002_35}
We used the periodic cell composed of six dimer lines along the transport normal direction (14.80 \AA) and sampled 66 $\vec{k_{\perp}}$ points. Along the transport direction, we used eight and two zigzag chains to model the channel and electrode regions, respectively, and the surface Green's functions were obtained for the corresponding electrode models sampled with the 25 $\vec{k_{\parallel}}$ points (see Supporting Information Fig. S1). In calculating $T(E)$, the energy was scanned from $-1.0$ eV to $1.0$ eV with respect to 
$E_F$ with the $0.001$ eV
resolution. In many cases, we have crosschecked the validity of our results using \textsf{SeqQuest} and our in-house NEGF code.
\cite{Y.-H.Kim_2005_36,Y.-H.Kim_2006_37}

\section{Results and Discussion\label{sec:results and discussion}}

\begin{figure*}[h]
\centering
 \includegraphics[scale=1.00]{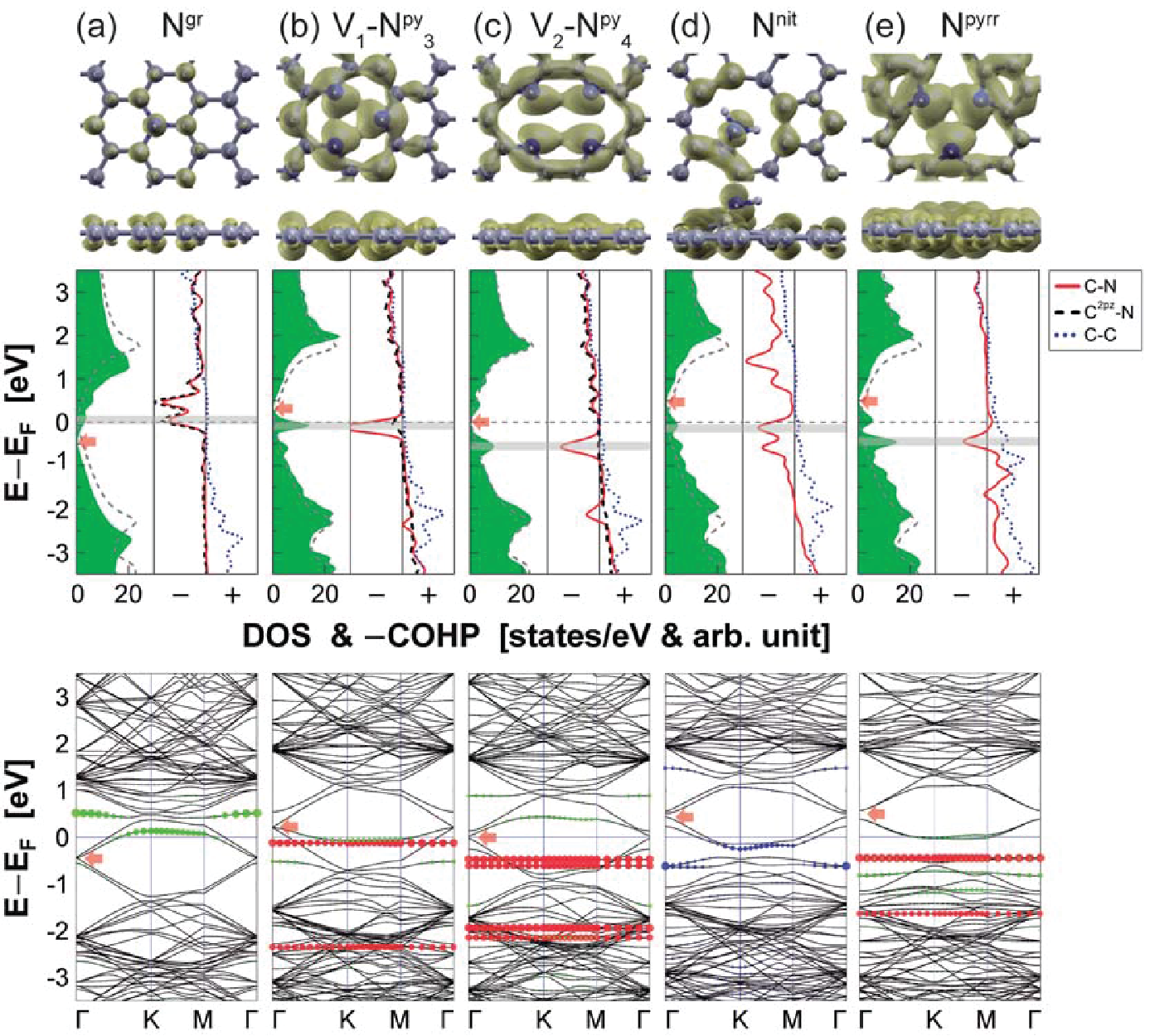}
  \caption{Atomic structures (top panels), DOS (middle left panels), COHPs (middle right panels), and bandstructures (bottom panels) of (a) N$^{gr}$, (b) V$_{1}$-N$^{py}_{3}$, (c) V$_{2}$-N$^{py}_{4}$, (d) N$^{nit}$, and (e) N$^{pyrr}$. On top of the atomic structures, we overlay the LDOS for the energy ranges marked as shaded regions in the DOS/COHP plots at the isovalues $5\times10^{-4} e\cdot$\AA$^{-3}$. In the DOS plots, pristine graphene DOS are shown together (gray dashed lines), and the Dirac points are marked by red arrows. In the COHP plots, we show the C-C (blue dotted lines) and C-N (red solid lines) bond curves, and additionally the C 2p$_{z}$ orbitals-N bond data (black dashed lines) for V$_{1}$-N$^{py}_{3}$ and V$_{2}$-N$^{py}_{4}$. In bandstuructures, green, red, and blue dots represent the projection of N $\pi$ states, N $\sigma$ states, and N atom, respectively. The weight is represented by dot size. \label{fig:Fig1}}
\end{figure*}

In this work, in addition to graphitic N (N$^{gr}$ (Fig.~\ref{fig:Fig1}a), we considered four other representative NG conformations: pyridinic N next to a monovacancy (V$_{1}$-N$^{py}$, 
Fig.~\ref{fig:Fig1}b) and a divacancy (V$_{2}$-N$^{py}$, 
Fig.~\ref{fig:Fig1}c), nitrilic N next to a divacancy (N$^{nit}$, 
Fig.~\ref{fig:Fig1}d), and pyrrolic N next to a trivacancy (N$^{pyrr}$, 
Fig.~\ref{fig:Fig1}e).
\cite{T.Schiros_2012_25,J.R.Pels_1995_38,I.Shimoyama_2000_39}
Recent experimental and theoretical reports showed that, despite high formation energies, vacancy defects can form within the graphene sheets by, $e.g.$, ion or electron irradiation.
\cite{F.Banhart_2010_40}
In the presence of vacancy defects, the pyridinic N configurations were shown to become energetically more favourable than 
the graphitic N one.
\cite{Y.Fujimoto_2012_26,Z.Hou_2012_41}
\begin{table}[h]
\small
  \caption{\ The formation energies of various N doped graphene models considered in this work.}
  \label{tbl:Table 1}
  \begin{tabular*}{0.5\textwidth}{@{\extracolsep{\fill}}ccccc}
    \hline
    \textrm{}&
    \multicolumn{4}{c}{\textrm{Formation energy [eV]}}\\
    \textrm{Number of N atoms}&\textrm{V$_{1}$-N$^{py}$}&\textrm{V$_{2}$-N$^{py}$}&\textrm{N$^{nit}$}&\textrm{N$^{pyrr}$}\\
    \hline
    1 & 5.14 	& 7.44 	& 7.66 	& -\\
    2 & 4.48 	& 5.53 	& - 		& -\\
    3 & 2.76 	& 4.77 	& - 		& 9.06\\
    4 & - 	& 2.90 	& - 		& -\\
    \hline
  \end{tabular*}
\end{table}

In Table 1, we show the zero-temperature formation energies of the V$_{1}$-N$^{py}$, V$_{2}$-N$^{py}$, N$^{nit}$, and N$^{pyrr}$ structures calculated according to
\begin{eqnarray}
 E_{f}(\textrm{NG}) = && (E_{N-doped} + x\mu_{C}) \nonumber \\*  
  	&& - ( E_{pristine}  + y\mu_{N} + z\mu_{H} ),
\end{eqnarray}
where E$_{N-doped/pristine}$ are the total energies of N-doped/pristine graphenes, $\mu_{C/N/H}$ are the chemical potentials (total energies per atom in their elemental reference phases) of C/N/H, $x$ is the number of C atoms removed from the graphene sheet during the vacancy formation, $y$ is the number of N atoms, and $z$ is the number of H atoms (2 for N$^{nit}$ and 0 otherwise).  The chemical potentials of C, N, and H were extracted from graphene, N$_2$ molecule, and H$_2$ molecule, respectively. 
Note that we include two pyridinic conformations, because the V$_{2}$-N$^{py}$ configuration is energetically even more favorable than the N$^{nit}$ and N$^{pyrr}$ counterparts. For the V$_{1}$-N$^{py}$ and V$_{2}$-N$^{py}$ cases, we additionally considered the substitution of different numbers of N atoms around the vacancy sites, V$_{1}$-N$^{py}_{\alpha}$ ($\alpha = 1\sim3$) and V$_{2}$-N$^{py}_{\beta}$ ($\beta = 1\sim4$), and found that the maximum concentration of pyridinic N atoms around the mono- and di-vacancies (trimerized V$_{1}$-N$^{py}_{3}$, tetramerized pyridine-like V$_{2}$-N$^{py}_{4}$) is energetically preferred (Table 1).
For V$_{1}$-N$^{py}_{\alpha}$ and V$_{2}$-N$^{py}_{\beta}$, we found that the maximum concentration of pyridinic N atoms around the mono- and di-vacancies (trimerized V$_{1}$-N$^{py}_{3}$, tetramerized pyridine-like V$_{2}$-N$^{py}_{4}$) is energetically preferred (Table 1).
\cite{Y.Fujimoto_2012_26}

An important objective of the present work is to provide the atomistic understanding on the nitrogen bonding nature in NG and its modification upon codoping. For this purpose, we will demonstrate that the combination of bandstructure, DOS, Mulliken population, and COHP
\cite{R.Dronskowski_1993_42}
provides a systematic and detailed information that guides the route toward the desired n-type or bipolar conversion. The COHP is defined by the DOS multiplied by the Hamiltonian of the corresponding element, and the negative COHP values ($-$COHP) give positive and minus signs for the bonding and antibonding states, respectively. The DOS, COHP normalised by the number of bonds, band structure data of N$^{gr}$, V$_{1}$-N$^{py}_{3}$, V$_{2}$-N$^{py}_{4}$, N$^{nit}$, and N$^{pyrr}$ are shown in Fig. 1, and their electronic structures are summarised in Table 2.
\begin{table}[h]
\small
  \caption{The C-N bond lengths, work functions, doping types, and doping concentrations of different NG configurations and their cooped cases. For B+V$_2$-N$^{py}_4$, two C-N bond lengths (for the B-attached and non-B-attached sides, see Fig. 4a) are given.}
  \label{tbl:Table 2}
  \begin{tabular*}{0.5\textwidth}{@{\extracolsep{\fill}}ccccc}
\hline
Structure			& $l_{C-N}$ [\AA]	& $\Phi$ [eV] & Doping type 	& $n$ [$10^{12}$ cm$^{-2}$]\\

\hline
 pristine 			& 1.43		& 4.23 	& - 			& - 		\\
				&			&		&			&		\\
 N$^{gr} $			& 1.42 		& 3.86 	& \textit{n} 		& 11.31 	\\
				&			&		&			&		\\
 V$_1$-N$^{py}_3$		& 1.35 		& 4.48 	& \textit{p} 		& $-9.31$	\\
 B+V$_1$-N$^{py}_3$	& 1.41 		& 3.84 	& \textit{n} 		& 16.03	\\
 Al+V$_1$-N$^{py}_3$	& 1.38 		& 3.86 	& \textit{n} 		& 15.57	\\
 Si+V$_1$-N$^{py}_3$	& 1.38 		& 3.84 	& \textit{n} 		& 12.82	\\
 P+V$_1$-N$^{py}_3$	& 1.40 		& 3.93 	& \textit{n} 		& 10.58	\\
				&			&		&			&		\\
 V$_2$-N$^{py}_4$		& 1.35 		& 4.32 	& - 			& -		\\
 B+V$_2$-N$^{py}_4$	& 1.42 / 1.39 	& 3.88 	& \textrm{n} 	& 10.52	\\
 Al+V$_2$-N$^{py}_4$	& 1.40 		& 3.80 	& \textrm{n} 	& 9.80	\\
 Si+ V$_2$-N$^{py}_4$	& 1.40 		& 3.86 	& \textrm{n} 	& 11.48	\\
 P+ V$_2$-N$^{py}_4$	& 1.39 		& 3.86 	& \textrm{n} 	& 5.56	\\
 				&			&		&			&		\\
N$^{nit}$			& 1.35 		& 4.53 	& \textit{p} 		& $-11.68$	\\
B+N$^{nit}$			& 1.46 		& 4.13 	& - 			& -		\\
Al+N$^{nit}$			& 1.49 		& 4.13 	& - 			& -		\\
Si+N$^{nit}$			& 1.47 		& 4.34 	& \textit{p}		& $-8.82$	\\
P+N$^{nit}$			& 1.47 		& 4.11 	& - 			& -		\\
				&			&		&			&		\\
N$^{pyrr}$			& 1.41 		& 4.61 	& \textit{p} 		& $-19.36$	\\
B+N$^{pyrr}$		& 1.51 		& 4.20 	& - 			& -		\\
Al+N$^{pyrr}$		& 1.45 		& 4.15 	& - 			& -		\\
Si+N$^{pyrr}$		& 1.47 		& 4.22 	& -			& -		\\
P+N$^{pyrr}$		& 1.48 		& 4.21 	& - 			& -		\\
\hline
  \end{tabular*}
\end{table}

\begin{figure*}[h]
\centering
  \includegraphics[scale=1.0]{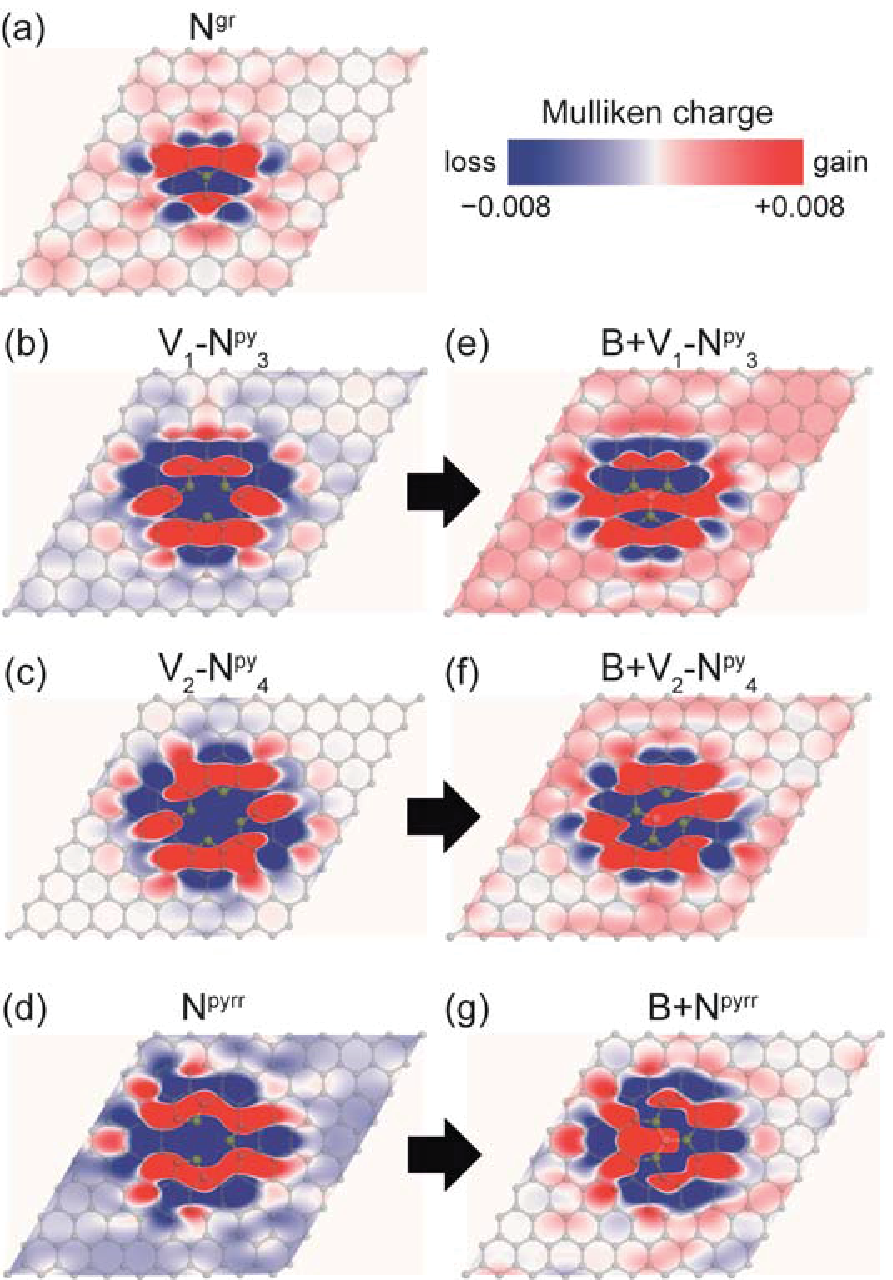}
  \caption{Mulliken population analysis of (a) N$^{gr}$, (b)V$_{1}$-N$^{py}_{3}$, (c) V$_{2}$-N$^{py}_{4}$, (d) N$^{pyrr}$, (e) B$+$V$_{1}$-N$^{py}_{3}$, (f) B$+$V$_{2}$-N$^{py}_{4}$, and (g) B$+$N$^{pyrr}$. The blue and red colors represent the loss and gain of electrons, respectively. \label{fig:Fig2}}
\end{figure*}

The doping type of graphitic NG is n-type, $i.e.$, $E_{D}$ is located below $E_{F}$ (Fig.~\ref{fig:Fig1}a middle and bottom panels). This results from the electron transfer from the N$^{gr}$ atoms to the $\pi$* states of graphene. Visualising the Mulliken charge populations as in Fig.~\ref{fig:Fig2}a, we can observe that the electrons donated from N$^{gr}$ are distributed throughout the graphene basal plane or become mobile charges. The amount of donated charges extracted from the population analysis is $0.402$ electron per nitrogen atom, which is in excellent agreement with the experimental estimation.
\cite{L.Zhao_2011_43}
It results in the upward shift of $E_{F}$ from $E_{D}$ of the pristine graphene case into the graphene conduction band, or the lowering of the work function by 0.37 eV. The calculated charge-carrier density of $11.31 \times 10^{12}$ cm$^{-2}$ for our model with a N atom doping ratio of 0.62 \% N atoms per C atom (N atom density of $2.31 \times 10^{13}$ cm$^{-2}$) is in good agreement with the experimental estimation of $5.42 \times 10^{12}$ electrons per cm$^2$ for the 0.34 \% N atoms per C atom doping. The COHP curve shows that these impurity resonant states have antibonding characters, being identified as the strong negative COHP peaks right above $E_{F}$ (thus $E_{D}$). The energetic locations of the antibonding COHP peaks are $E_{F} + 0.07$ eV and $E_F + 0.45$ eV. These peak locations depend on the supercell size, or the doping ratio, and is expected to converge to an experimentally observed single peak at about $E_{F} +0.14$ eV.
\cite{Z.Hou_2013_27,T.Kondo_2012_44}

On the other hand, the electronic structures of the other three V$_{1}$-N$^{py}_{3}$, N$^{nit}$, and N$^{pyrr}$ NG are p-type, or $E_{F}$ has been shifted downward into the valence bands (Figs.~\ref{fig:Fig1}b, ~\ref{fig:Fig1}d, and~\ref{fig:Fig1}e middle and bottom panels). This should mainly result from the presence of vacancy defect states (whose LDOS are shown in Figs.~\ref{fig:Fig1}b,~\ref{fig:Fig1}d, and~\ref{fig:Fig1}e top panels), which behave as acceptor states with missing $\pi$-electrons.
\cite{D.Usachov_2011_24,T.Schiros_2012_25,Z.Hou_2013_27}
Note that in these cases the impurity states appear in the COHP plots as strong antibond peaks right below $E_{F}$ (thus $E_{D}$). Interestingly, the V$_{2}$-N$^{py}_{4}$ case shows a bipolar character ($E_{D}$ nearly coincides with $E_{F}$).
\cite{Z.Hou_2013_27}
To understand such discrepancies, we analyzed Mulliken populations as shown in Figs.~\ref{fig:Fig2}b, \ref{fig:Fig2}c, and \ref{fig:Fig2}d for V$_{1}$-N$^{py}_{3}$, V$_2$-N$^{py}_4$ and N$^{pyrr}$, respectively (The N$^{nit}$ diagram is similar to N$^{pyrr}$ and is not shown). They show that the spatial range of charge redistribution around the nitrogen-vacancy complexes discriminate the V$_{1}$-N$^{py}_{3}$ and N$^{pyrr}$ cases from the V$_{2}$-N$^{py}_{4}$ counterpart: We observe the depletion of electrons throughout the entire graphene basal plane in the former, but rather localised charge depletion around the nitrogen-vacancy site in the latter (Compare Fig.~\ref{fig:Fig2}c with Figs.~\ref{fig:Fig2}b and \ref{fig:Fig2}d). Related with these charge transfer characters, the impurity states strongly hybridize with the carbon 2p$_{z}$ orbitals or graphene $\pi$ states (existence of an antibonding C 2p$_{z}$-N COHP peak) in V$_{1}$-N$^{py}_{3}$ (Fig. \ref{fig:Fig1}b COHP panel, shaded region near $E_F$), whereas the hybridisation is very weak (absence of an antibonding C 2p$_{z}$-N COHP peak) in V$_{2}$-N$^{py}_{4}$ (Fig. \ref{fig:Fig1}c COHP panel, shaded region at $E - E_F \approx$ 0.5 eV).
\begin{figure*}[h]
\centering
  \includegraphics[scale=1.0]{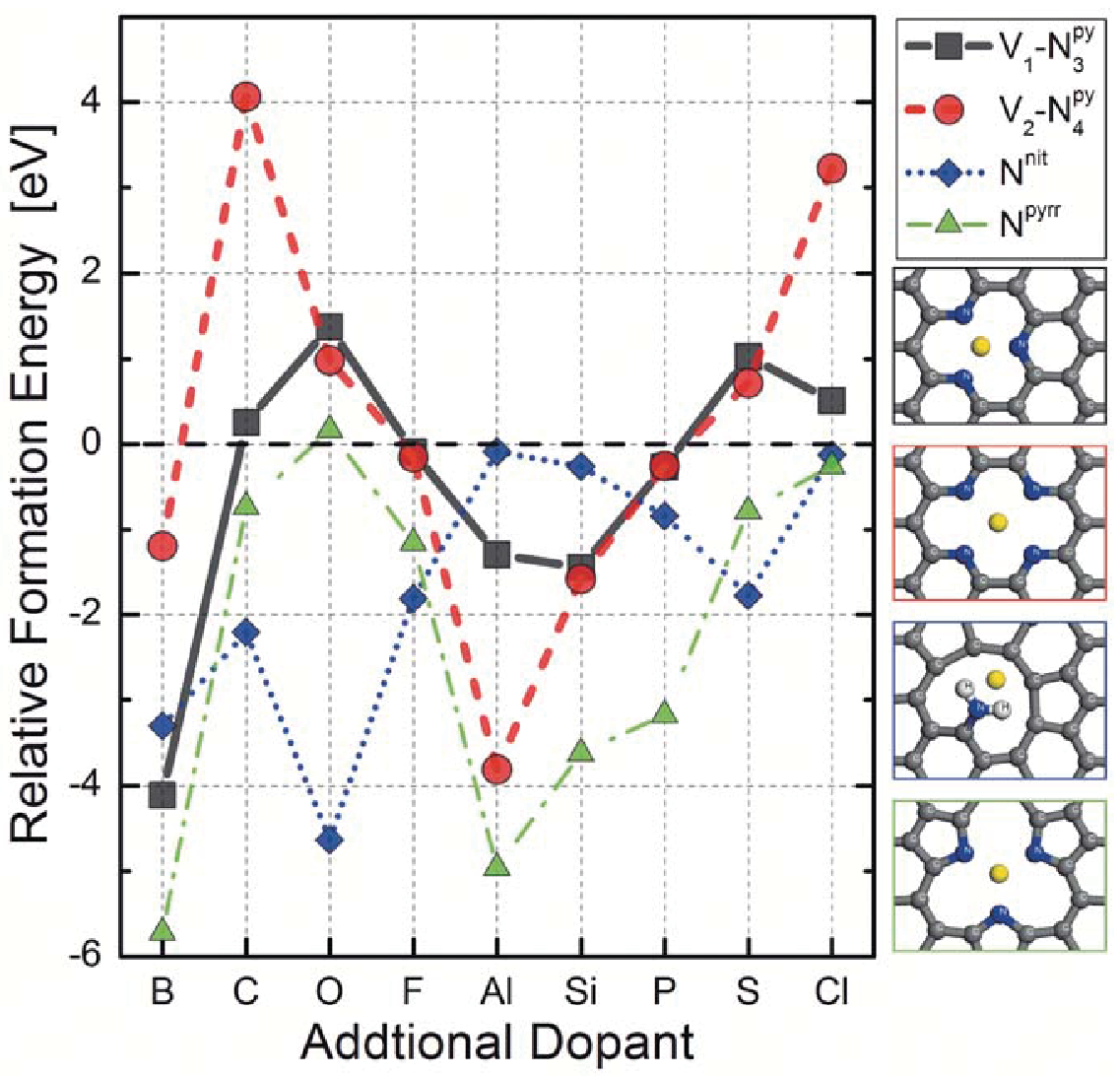}
  \caption{Relative formation energies of introducing an additional dopant atom into V$_{1}$-N$^{py}_{3}$ (black squares), V$_{2}$-N$^{py}_{4}$ (red circles), N$^{nit}$ (blue diamonds), and N$^{pyrr}$ (green triangles). \label{fig:Fig3}}
\end{figure*}

We now move on to consider the possibility of codoping the p-type NG. Since the presence of a vacancy defect is the precondition of the existence of V$_{1}$-N$^{py}_{3}$, V$_{2}$-N$^{py}_{4}$, N$^{nit}$, and N$^{pyrr}$ sites, we focused on whether the vacancy defects can be healed by introducing a codopant. In addition to C, eight light elements, B, O, F, Al, Si, P, S, and Cl, were considered as codoping candidates. For the fully optimised structures, $X+$V$_{1}$-N$^{py}_{1\sim3}$, $X+$V$_{2}$-N$^{py}_{1\sim4}$, $X+$N$^{nit}$, and $X+$N$^{pyrr}$ ($X =$ B, C, O, F, Al, Si, P, S, and Cl), we have calculated the formation energies,
%
\begin{eqnarray}
 E_{f}(X\textrm{+NG}) = && (E_{codoped} + x\mu_{C}) \nonumber \\* 
 		&& - (E_{pristine}  + y\mu_{N} + z\mu_{H} + \mu_X),
\end{eqnarray}
where $E_{codoped}$ 
is the total energy of codoped NG, $\mu_{C/N/H/X}$ are the chemical potentials of C, N, H, and the codopant $X$, $x/y$ are the numbers of removed/added C/N atoms with respect to the pristine graphene, and $z$ is the number of added H atoms (2 for N$^{nit}$ and 0 otherwise). The chemical potentials of B, O, F, Al, Si, P, S, and Cl were obtained from the $\alpha$-rhombohedral boron, O$_2$ molecule, F$_2$ molecule, face-centred cubic aluminum, diamond cubic Si, black phosphorus, orthorhombic sulphur ($\alpha$-S$_8$), and Cl$_2$ molecule, respectively.
The formation energies relative to those of NG,
\begin{equation}
 E_{rf}(X\textrm{+NG}) = E_{f}(X\textrm{+NG}) - E_{f}(\textrm{NG}),
\end{equation}
which represent the energetic feasibility of annealing the vacancy defects of p- or bipolar-type NG by the codoping approach, are summarised in Fig.~\ref{fig:Fig3} (For V$_{1}$-N$^{py}_{1\sim3}$ and V$_{2}$-N$^{py}_{1\sim4}$, only the energetically most favourable V$_{1}$-N$^{py}_{3}$ and V$_{2}$-N$^{py}_{4}$ cases are shown. See Supporting Information Figs. S2 and S3 for other cases).
\begin{figure*}[h]
\centering
  \includegraphics[scale=1.2]{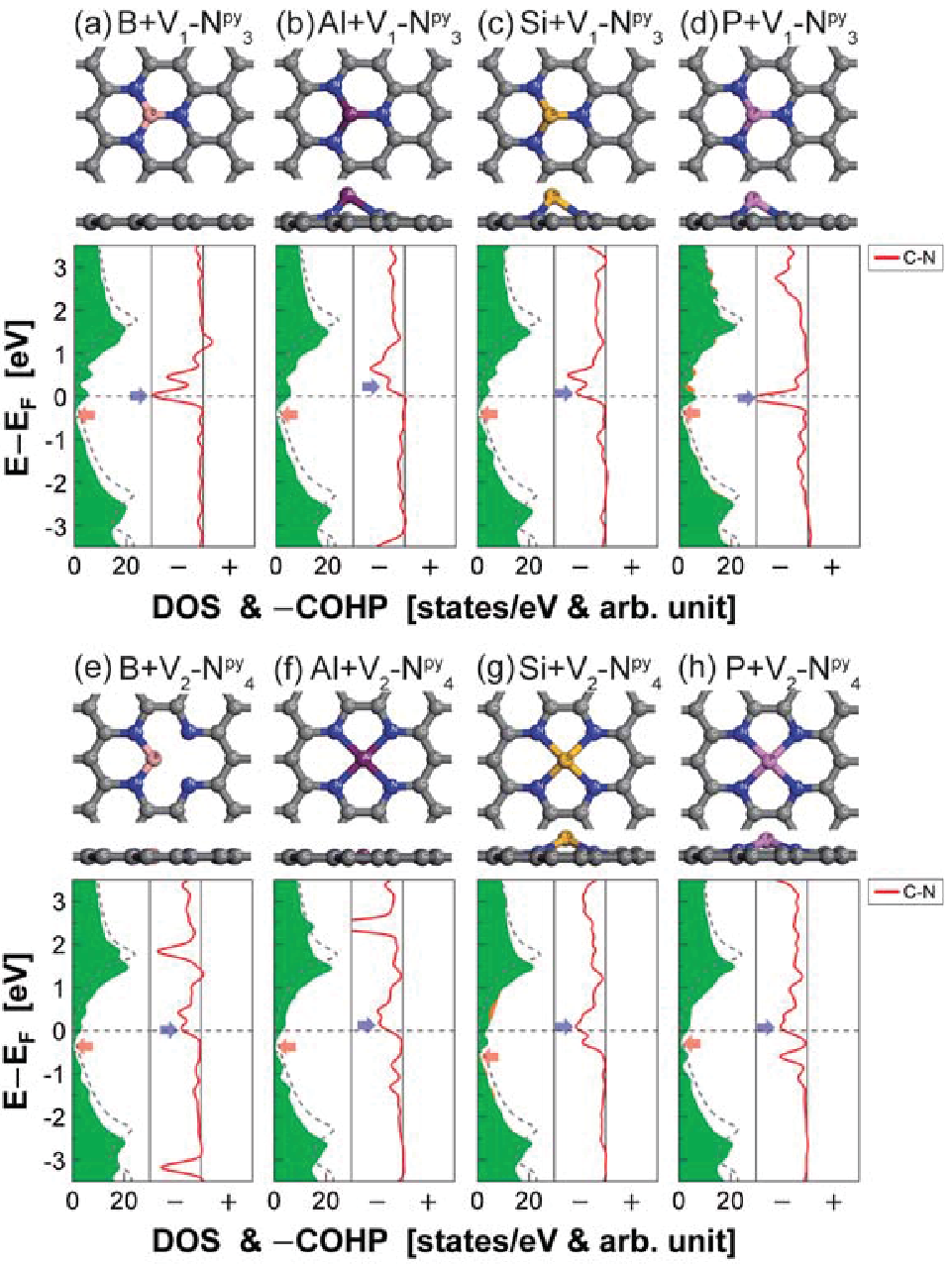}
  \caption{Atomic structures (top panels), DOS (bottom left panels), and COHPs (bottom right panels) of (a) B$+$V$_{1}$-N$^{py}_{3}$, (b) Al$+$V$_{1}$-N$^{py}_{3}$, (c) Si$+$V$_{1}$-N$^{py}_{3}$, (d) P$+$V$_{1}$-N$^{py}_{3}$, (e) B$+$V$_{2}$-N$^{py}_{4}$, (f) Al$+$V$_{2}$-N$^{py}_{4}$, (g) Si$+$V$_{2}$-N$^{py}_{4}$, and (h) P$+$V$_{2}$-N$^{py}_{4}$. In the DOS plots, pristine graphene DOS are shown together (gray dashed lines), and the Dirac points are marked by red arrows. In the COHP plots, we show the C-N (red solid lines) bond data. For the slightly spin-polarized P$+$V$_{1}$-N$^{py}_{3}$ and P$+$V$_{1}$-N$^{py}_{3}$ cases, we show both the majority (green filled lines) and minority (orange filled lines) DOS. \label{fig:Fig4}}
\end{figure*}

First, we find that, except for O$+$N$^{pyrr}$, both N$^{nit}$ and N$^{pyrr}$ defects can be annealed by all of the above codoping elements (negative $E_{rf}$), because the formation energies of N$^{nit}$ and N$^{pyrr}$ are rather large to begin with (Table 1) and their relatively large vacancies can easily accommodate an additional dopant atom. However, incorporating an additional dopant atom into the energetically more favourable pyridinic NG is found to be rather difficult: It was determined that the V$_{1}$-N$^{py}_{3}$ and V$_{2}$-N$^{py}_{4}$ defects can be annealed only with a B, F, Al, Si, or P atom. It was determined that the energetically less favourable pyridinic NG with lower N concentrations can also accommodate these elements (see Supporting Information Figs. S2 and S3). The results, $e.g.$, indicates that a N$^{py}$ to N$^{gr}$ conversion scheme solely based on the carbon source
\cite{D.Usachov_2011_24}
will be limited in that di- or larger vacancies cannot be annealed. Although the F codoping of p-type NG might be energetically feasible, it is found that the F atom cannot structurally passivate the vacancy defects in the V$_{1}$-N$^{py}_{3}$ and V$_{2}$-N$^{py}_{4}$ cases and accordingly the electronic structures of F$+$V$_{1}$-N$^{py}_{3}$ and F$+$V$_{2}$-N$^{py}_{4}$ still maintain the p-type and bipolar characters, respectively (see Supporting Information Fig. S4). This leaves B, Al, Si, and P as the codoping candidates for the n-type conversion of p-type NG.

We first focus on the pyridinic NG codoped with B, Al, Si, and P, whose fully relaxed geometries and electronic structures are shown in Fig.~\ref{fig:Fig4}. The C-N bond lengths of pyridinic NG are elongated when the vacancy is annealed by the codopant (Table 2). We further find that the trend of energetic stability of codoped NG (Fig.~\ref{fig:Fig3}) is strongly related with the planarity of their geometries. For example, B can anneal the V$_{1}$-N$^{py}_{3}$ defect without protrusion (Fig.~\ref{fig:Fig4}a), which makes its $E_{rf}$ larger than those of Al, Si, and P. Similarly, Al can anneal the V$_{2}$-N$^{py}_{4}$ defect in a fourfold-coordinated planar configuration (Fig. \ref{fig:Fig4}f), which makes it an energetically more favourable codopant than B, Si, and P. 

In terms of electronic structure (Fig.~\ref{fig:Fig4} bottom panels and Table 2), we find that all the $X+$V$_{1}$-N$^{py}_{3}$ and $X+$V$_{2}$-N$^{py}_{4}$ cases ($X$ = B/Al/Si/P) become n-type, or $E_{D}$ is now located below $E_{F}$.  The passivation of V$_{1}$-N$^{py}_{3}$ and V$_{2}$-N$^{py}_{4}$ vacancy defects by the codoping of B/Al/Si/P allows the charge transfer from the N atoms to the graphene $\pi$* states (Figs.~\ref{fig:Fig2}e and \ref{fig:Fig2}f) as in the case of N$^{gr}$ (Fig.~\ref{fig:Fig2}a). The alteration of the electronic type change is also easily identified in the COHP data, in which we observe strong antibonding C-N peaks that are pinned near $E_{F}$ or the downward shift of $E_{D}$ below $E_{F}$.  The reduction of work function for V$_1$-N$^{py}_3$ and V$_2$-N$^{py}_4$ was up to 0.62 eV and 0.52 eV upon codoping of B/Si and Al, respectively.

To further understand the microscopic mechanisms of electronic type conversion in $X+$V$_{1}$-N$^{py}_{3}$ and$X+$V$_{2}$-N$^{py}_{4}$, we have decomposed the C-N COHPs into different orbital contributions (Fig. S5) and found that the C-N antibonding COHP peaks mostly come from the C 2p$_{z}$-N antibonds. Namely, the insertion of a B/Al/Si/P atom into V$_{1}$-N$^{py}_{3}$ or V$_{2}$-N$^{py}_{4}$ and the subsequent  annealing of vacancy allows the N atoms to directly couple with the sp$^{2}$ C network and behave like the graphitic Ns. Whereas the amount of downward shift of $E_{D}$ tends to decrease with the lowering of N concentrations, we have confirmed that the B, Al, Si, and P codoping of V$_{1}$-N$^{py}_{1\sim2}$ and V$_{2}$-N$^{py}_{1\sim3}$ in general changes their electronic type to n- or at least bipolar ones (see Supporting Information Figs. S6-S10).
\begin{figure*}[h]
\centering
  \includegraphics[scale=1.2]{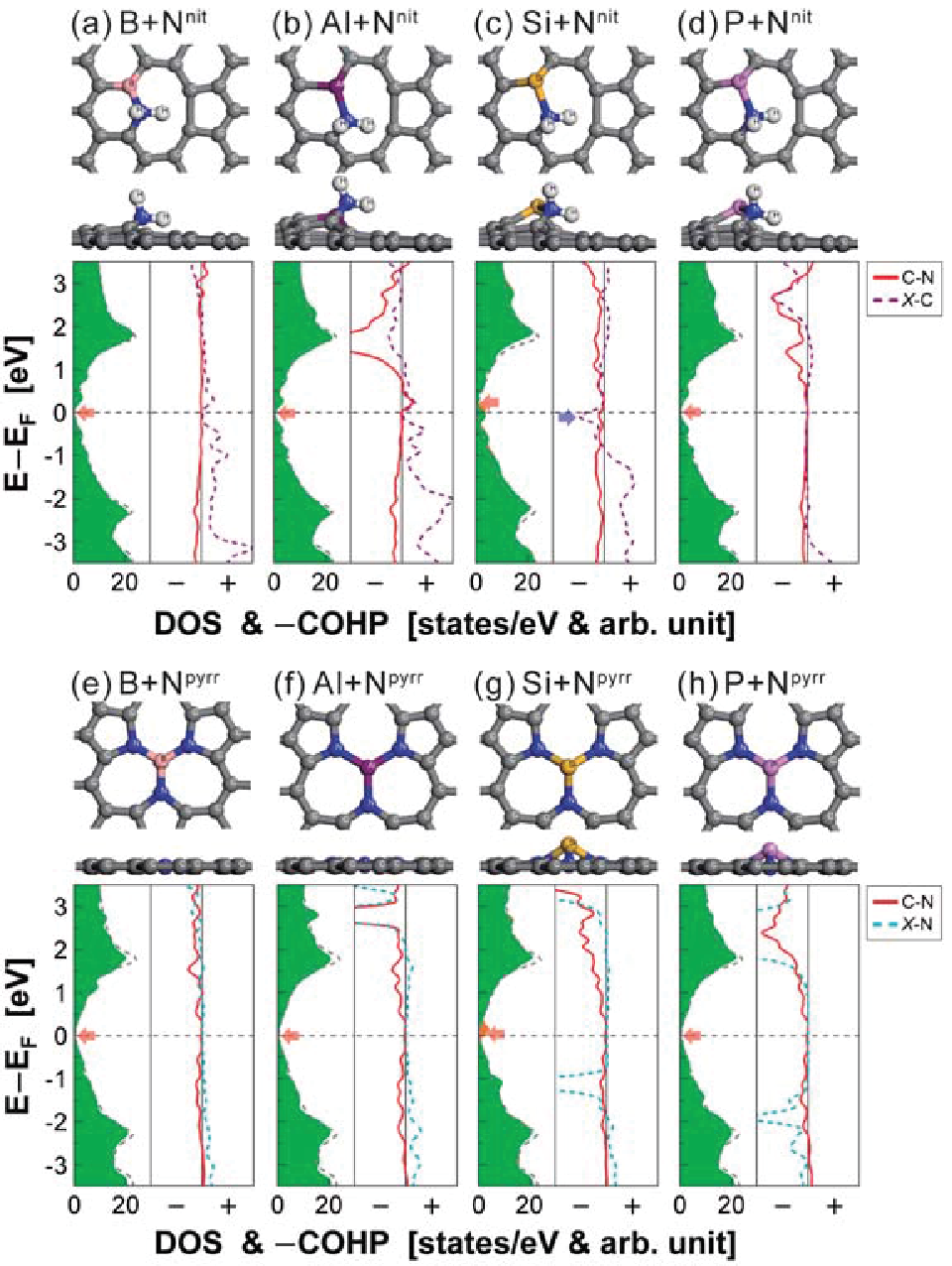}
  \caption{Atomic structures (top panels), DOS (bottom left panels), and COHPs (bottom right panels) of (a) B$+$N$^{nit}$, (b) Al$+$N$^{nit}$, (c) Si$+$N$^{nit}$, (d) P$+$N$^{nit}$, (e) B+Npyrr, (f) Al$+$N$^{pyrr}$, (g) Si$+$N$^{pyrr}$, and (h) P$+$N$^{pyrr}$. In the DOS plots, pristine graphene DOS are shown together (gray dashed lines), and the Dirac points are marked by red arrows. In the COHP plots, the C-N (red solid lines), $X$-C (purple dashed lines), and $X$-N (cyan dashed lines) bond curves are shown. For the slightly spin-polarized Si$+$N$^{nit}$ and Si$+$N$^{pyrr}$ cases, we show both the majority (green filled lines) and minority (orange filled lines) DOS. \label{fig:Fig5}}
\end{figure*}

Compared with the pyridinic cases, the nitrilic and pyrrolic NG show a different type conversion behavior and thus become interesting comparative systems (Fig.~\ref{fig:Fig5} and Table 2). The C-N bond lengths in nitrilic and pyrrolic NG are more elongated than in the pyridinic counterparts. We also observe that, except the case of Si$+$N$^{nit}$ (Fig.~\ref{fig:Fig5}c bottom panels), all of them show the bipolar property (Table 2). The different behaviour between the N$^{py}$ family (can be converted to the n-type) and N$^{nit}$ or N$^{pyrr}$ (can be converted only to the bipolar-type) upon the introduction of a codoping element can be understood from their band structure plots. While the pyridinic defect states can energetically mix with the $\pi$ states of graphene (Figs.~\ref{fig:Fig1}b and \ref{fig:Fig1}c bottom panels), the nitrilic and pyrrolic defects states are energetically located further away from $E_{F}$ and cannot strongly hybridise with the graphene $\pi$ states 
(Figs.~\ref{fig:Fig1}d and \ref{fig:Fig1}e bottom panels).

In spite of the difference, we can generally conclude that the p-type character of the structurally more distorted and energetically less probable N$^{nit}$ and N$^{pyrr}$ defects (Table 1) can be also eliminated by the B, Al, and P codoping. The COHPs of C-N in these cases are nearly zero around $E_{D}$, or the N atoms in the codoped N$^{nit}$ and N$^{pyrr}$ do not affect the sp$_{2}$ carbon network. The nature of charge distribution in $X+$N$^{nit}$ and $X+$N$^{pyrr}$ can be again visualised in the Mulliken population analysis plot as shown for the representative B$+$N$^{pyrr}$ case in Fig. 2g, which shows that the net induced charge is localised near the defect sites as in the (bipolar) V$_{2}$-N$^{py}_{4}$ (Fig. 2c).
\begin{figure*}[h]
\centering
  \includegraphics[scale=1.4]{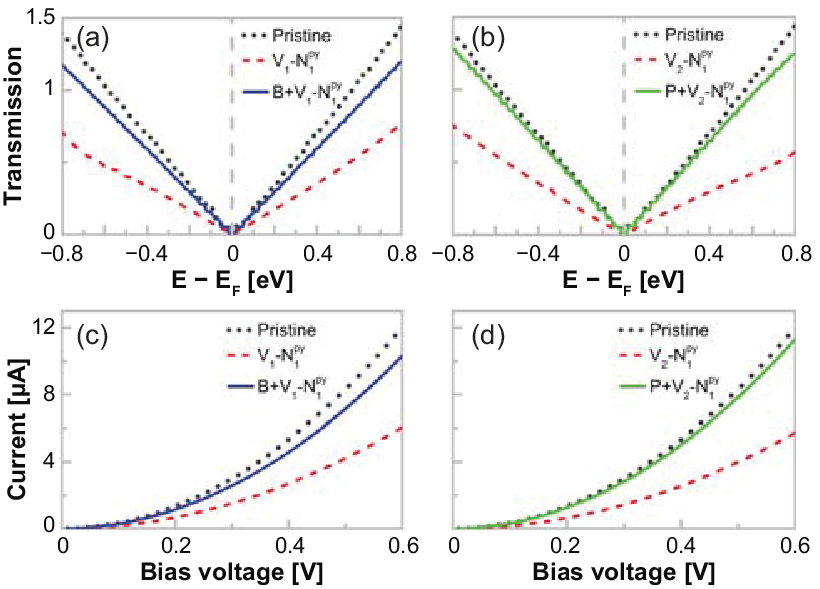}
  \caption{Transmission functions of (a) V$_{1}$-N$^{py}_{1}$ (red dashed line) and B$+$V$_{1}$-N$^{py}_{1}$ (blue solid line), and (b) V$_{2}$-N$^{py}_{1}$ (red dashed line) and P$+$V$_{2}$-N$^{py}_{1}$ (green solid line). The pristine graphene transmissions are shown together (grey dotted lines). \label{fig:Fig6}}
\end{figure*}

It remains to explain why the Si-codoped N$^{nit}$ maintains the p-type character (Fig. 5c bottom panel). To understand its origin, we have additionally analysed the COHPs of various bonds (See Supporting Information Fig. S11), and found that Si$+$N$^{nit}$ can be discriminated from the B/Al/P$+$N$^{nit}$ counterparts by a strong antibonding Si-C COHP peak that appears below $E_{D}$ and is pinned near $E_{F}$ (Fig.~\ref{fig:Fig5}c bottom panel). On the other hand, $e.g.$, the B-C in B$+$N$^{nit}$ has the bonding character around $E_{F}$ (Fig.~\ref{fig:Fig5}a bottom panel).

We finally demonstrate a very desirable feature of the codoping approach in terms of charge transport characteristics. We show in Figs.~\ref{fig:Fig6}a and ~\ref{fig:Fig6}b the transmission functions of (codoped) NG for the representative cases of (B$+$)V$_{1}$-N$^{py}_{1}$ and (P$+$)V$_{2}$-N$^{py}_{1}$. 
Note that in these NEGF calculations, unlike the DFT results that have been discussed so far, we are considering isolated doping sites sandwiched by two semi-infinite \textit{pristine} graphene electrodes (See Supporting Information Fig. S1).  The work function of the entire junction system is thus determined by the pristine graphene region, which results in the Dirac point located at $E_F$.
The resulting current-bias voltage curves calculated according to the Landauer-B\"uttiker formula,
\begin{equation}
 I(V) = \frac{2e}{h} \int_{\mu_1}^{\mu_2} dE \, T(E)[f(E-\mu_1)-f(E-\mu_2)],
\end{equation}
where $\mu_1 = E_F - 0.5 eV$ and $\mu_2 = E_F + 0.5 eV$, are shown together in Figs.~\ref{fig:Fig6}c and ~\ref{fig:Fig6}d, respectively. For the untreated p-type NG models (V$_{1}$-N$^{py}_{1}$ and V$_{2}$-N$^{py}_{4}$), we find that the localised defect states around vacancy sites result in currents much decreased from that of pristine graphene. However, upon introducing codopant atoms and healing the vacancy defects (B$+$V$_{1}$-N$^{py}_{1}$ and P$+$V$_{2}$-N$^{py}_{1}$), we almost completely recover the currents of pristine graphene. 
We have recently obtained a similar behaviour for the B-N edge-doped graphene nanoribbons
\cite{S.S.Kim_2014}.

The availability of a simple yet effective method to convert NG with mixed n- and p-type characters into near-uniform n-type  
NG will improve the performance and reliability of various graphene-based electronic devices such as all-graphene p-n junctions, field-effect transistors, integrated circuits, $etc$.
\cite{Y.-M.Lin_2011_45,N.M.Gabor_2011_46,F.Withers_2013_47,F.Al-Dirini_2011_48}
We additionally note that the codoping-induced type change in NG may have significant implications for energy applications.
\cite{U.N.Maiti_2014_10,Y.Wang_2010_11,L.Qu_2010_12,A.L.M.Reddy_2010_13,H.Wang_2011_14,Z.Luo_2011_15,H.M.Jeong_2011_16,L.Lai_2012_17,H.Wang_2012_18,M.J.Ju_2013_19,X.Wang_2014_20}
In the effort to identify the catalytically active sites in NG that result in significantly improved ORR performance, it was recently shown that the ORR activity in NG-based fuel cell cathodes is proportional to the graphitic N content.
\cite{L.Lai_2012_17}
On the other hand, it was experimentally demonstrated that the codoping of B or P into NG improves the ORR performance.
\cite{S.Wang_2012_28,C.H.Choi_2013_29}
While the precise origin of the enhanced ORR activity in NG and codoped NG is not clear yet, we can propose that the codoping-induced conversion of mixed-type NG into uniformly n-type NG ($i.e.$ generating more ``effectively'' graphitic N atoms) and the enhanced charge transport capacity will play an important role in enhancing the ORR activity.

\section{Conclusion\label{sec:conclusion}}

In summary, based on fully first-principles computations, we have predicted that the post-synthetic codoping of p-type NG with B, Al, and P can anneal even relatively large vacancy defects associated with pyridinic, pyrrolic, and nitrilic Ns. We also found that, while the pyridinic NG can be converted into n-type, only the bipolar type conversion is allowed for the nitrilic and pyrrolic counterparts. More importantly, employing novel analysis schemes, we have revealed the origin of p-type or bipolar behaviour of NG, and showed that the codoping and consequent healing of vacancy defects strongly modify the C-N antibonding character, its energetic position, and the spatial distribution of electrons donated from N atoms. We expect this will provide in the future a framework to systematically characterise doped and/or defective graphene. Furthermore, we demonstrated that the codoping of p-type NG can recover the excellent charge transport capacity of pristine graphene, and suggested that together with the electronic type change it should have significant implications for energy and electronic device applications. In view of experimental advances already achieved in the control of doping characters of NG,
\cite{Z.Luo_2011_15,L.Lai_2012_17}
we envisage the facile post-synthetic codoping scheme proposed here will lead to new pathways for tailoring and enhancing the properties of graphene at the atomic level.

\textit{Note added in proof}: After submission of this work, we became aware of Ref. 
\citenum{Y.Xue_2014},
in which the authors experimentally synthesised P, N-codoped graphene by chemical vapour deposition method and found much improved air-stable n-type characteristics.
In addition, in Ref. 
\citenum{Y.Fujimoto_2014},
the authors theoretically predicted that the dissociative adsorption of H$_2$ molecule on the trimerized and tetramerized pyridine-type defects is energetically favourable, and the adsorption of the two H atoms change their electronic properties from p-type to n-type doping.

\begin{acknowledgements}
This research was supported mainly by Global Frontier Program (2013M3A6B1078881), and additionally by Basic Science Research Grant (No. 2012R1A1A2044793), Nano$\cdot$Material Technology Development Program (No. 2012M3A7B4049888), and EDISON Program (No. 2012M3C1A6035684) of the National Research Foundation funded by the Ministry of Education Science and Technology of Korea. Computational resources were provided by the KISTI Supercomputing Center (No. KSC-2013-C3-046)
\end{acknowledgements}

\nocite{*}


%

\end{document}